# Comment on "Observation of a push force on the end face of a nanometer silica filament exerted by outgoing light," *Phys. Rev. Lett.* **101,** 243601 (2008)

Masud Mansuripur

College of Optical Sciences, The University of Arizona, Tucson, Arizona 85721

In a recent paper [1], W. She, J. Yu and R. Feng reported the slight deformations observed upon transmission of a light pulse through a fairly short length of a silica glass nano-fiber. Relating the shape and magnitude of these deformations to the momentum of the light pulse both inside and outside the fiber, these authors concluded that, within the fiber, the photons carry the Abraham momentum. In my view, the authors' claim that they have resolved the Abraham-Minkowski controversy surrounding the momentum of photons inside dielectric media is premature. A correct interpretation of the experiments of She *et al* requires precise calculations that would properly account not only for the electromagnetic momentum (both inside and outside the fiber) but also for the Lorentz force exerted on the fiber by the light pulse in its *entire* path through this nano-waveguide [2-4]. My comments below focus only on the theoretical interpretation of what She *et al* have observed. There are serious issues with their experimental method as well, but space limitations prevent me from describing them here in any detail.

1) The momentum of a light pulse in vacuum is given by $\mathcal{E}_{\text{pulse}}/c$ – the ratio of the pulse energy to the speed of light in vacuum – only when the pulse has a cross-sectional diameter (in the *xy*-plane perpendicular to the propagation direction *z*) that is much greater than the light's wavelength $\lambda$. In other words, $\boldsymbol{p} = (\mathcal{E}_{\text{pulse}}/c)\hat{\boldsymbol{z}}$ is valid only when there is negligible diffraction-broadening during propagation along *z*. The light spots emerging from the fiber in Figs. 1, 2 and 4 of [1] are ~500 nm in diameter, comparable to a wavelength ($\lambda_1 = 650$ nm, $\lambda_2 = 980$ nm). Consequently, the emergent momentum along *z* is much less than $\mathcal{E}_{\text{pulse}}/c$, resulting in a substantial error in the formulas listed in [1], page 1, 2$^{\text{nd}}$ column, lines 1-12.

In the original Einstein box Gedanken experiment (empty box on a frictionless rail, light emerging from the wall on the left, traveling the length of the box, impinging on a perfect absorber on the right-hand side), one must recognize that, if the pulse at the left wall has a small cross-section (e.g., emanating from a nano-fiber), it will expand, due to diffraction, into a spherical wave as it propagates to the right. The smaller the transverse dimensions of the light source, the greater will be the deviation of the pulse's momentum from $\mathcal{E}_{\text{pulse}}/c$. Assuming an Einstein box with a point-dipole oscillator at the center of a hemi-spherical absorbing surface, the pulse momentum can be shown to be $\boldsymbol{p} = (3\mathcal{E}_{\text{pulse}}/4c)\hat{\boldsymbol{z}}$. The light exiting the fiber thus carries only 75% of the momentum assigned to it by She *et al*.

2) The refractive index *n* that defines the Abraham momentum inside a dielectric is the *group* index, not the *phase* index used by She *et al*. In an ordinary glass slab, one might be able to ignore dispersion effects and assume dispersionless propagation, in which case the group and phase indices will be nearly the same. The validity of this "dispersionless" approximation, however, is no longer obvious when dealing with propagation inside a waveguide, especially when the silica filament has been pulled so drastically as to force a substantial fraction of the optical energy into the air surrounding the nano-fiber. By using in their calculations the bulk index of silica glass ($n = 1.451$ at $\lambda_1$, 1.457 at $\lambda_2$) – thus failing to account for the group velocity inside an extremely narrow waveguide – the authors have cast serious doubts on the validity of their interpretations.

3) The Abraham momentum is only one component of the momentum of light inside a dielectric, namely, the electromagnetic component $p_{EM}$; the other component, denoted by $p_{mech}$, is mechanical. ($p_{mech}$ is *not* the same thing as the difference between the Minkowski and Abraham momenta, as some have suggested; it is half as much under certain circumstances [4].) A correct accounting for the observed deformation of the nano-fiber would have required a complete balancing of the momenta, namely, $p_{EM}+p_{mech}$ inside the fiber, minus the pure electromagnetic momentum outside, where the light emerges into the free space. She *et al* have completely ignored the role of $p_{mech}$ inside their nano-fibers.